\documentclass[
reprint,
amsmath,
amssymb,
aps,
prd,
10pt,
tightenlines,
superscriptaddress,
nofootinbib,
preprintnumbers,
twocolumn,
notitlepage]{revtex4-2}

\usepackage{lineno}
\usepackage{amsmath,amssymb,amsthm,amsfonts}
\usepackage{graphicx,tabularx}
\usepackage{color}
\usepackage{multirow}
\usepackage{comment}
\usepackage{enumitem}
\usepackage{subfigure}
\usepackage{orcidlink}
\usepackage{cleveref}
\usepackage{slashed}
\usepackage[normalem]{ulem}
\usepackage{scalerel}
\usepackage{wrapfig}
\usepackage{cancel}
\usepackage{xcolor}
\usepackage{multirow}
\usepackage{setspace}
\usepackage{xspace}

\newcommand{\XSeventeen}{X17\xspace}

\newcommand{\PRE}[1]{{#1}} 

\newcommand{\be}{\begin{equation} \begin{aligned}}
\newcommand{\ee}{\end{aligned} \end{equation}}
\newcommand{\beqa}{\begin{eqnarray}}
\newcommand{\eeqa}{\end{eqnarray}}

\def\figureautorefname~#1\null{Fig.\,#1\null}
\def\tableautorefname~#1\null{Tab.\,#1\null}
\def\equationautorefname~#1\null{Eq.\,(#1)\null}

\newcommand{\mev}{\text{MeV}}
\newcommand{\gev}{\text{GeV}}
\newcommand{\tev}{\text{TeV}}

\newcommand{\ifb}{\text{fb}^{-1}}

\newcommand{\micm}{\mu\text{m}}
\newcommand{\mm}{\text{mm}}
\newcommand{\cm}{\text{cm}}
\newcommand{\m}{\text{m}}

\newcommand{\micrad}{\mu\text{rad}}

\RequirePackage[normalem]{ulem}

\crefname{section}{Sec.}{Secs.}
\crefname{figure}{Fig.}{Figs.}
\crefname{equation}{Eq.}{Eqs.}
\crefname{table}{Table}{Tables}
\crefname{appendix}{Appendix}{Appendices}

\newcommand{\blankfootnote}[1]{%
  \let\svthefootnote\thefootnote%
  \let\thefootnote\relax\footnotetext{#1}%
  \let\thefootnote\svthefootnote%
}

\makeatletter
\renewcommand{\p@subsection}{}
\renewcommand{\p@subsubsection}{}
\makeatother

\begin{document}

\preprint{CERN-EP-2026-216}
\preprint{2 November 2026}
\title{
\vspace*{0.2in}
{\Large Constraints on a Long-Lived \XSeventeen Vector Boson with FASER} 
\bigskip \\  
FASER Collaboration
}

\author{Roshan Mammen Abraham\,\orcidlink{0000-0003-4678-3808}}
\affiliation{Department of Physics and Astronomy, University of California, Irvine, CA 92697-4575, USA}

\author{Xiaocong Ai\,\orcidlink{0000-0003-3856-2415}}
\affiliation{School of Physics, Zhengzhou University, Zhengzhou 450001, China}
  
\author{Saul Alonso Monsalve\,\orcidlink{0000-0002-9678-7121}}
\affiliation{Institute for Particle Physics, ETH Z\"urich, Z\"urich 8093, Switzerland}

\author{John Anders\,\orcidlink{0000-0002-1846-0262}}
\affiliation{University of Liverpool, Liverpool L69 3BX, United Kingdom}

\author{Emma Kate Anderson\,\orcidlink{0000-0002-0161-4560}}
\affiliation{CERN, CH-1211 Geneva 23, Switzerland}

\author{Akitaka Ariga\,\orcidlink{0000-0002-6832-2466}}
\affiliation{Albert Einstein Center for Fundamental Physics, Laboratory for High Energy Physics, University of Bern, Sidlerstrasse 5, CH-3012 Bern, Switzerland}
\affiliation{Department of Physics, Chiba University, 1-33 Yayoi-cho Inage-ku, 263-8522 Chiba, Japan}

\author{Tomoko Ariga\,\orcidlink{0000-0001-9880-3562}}
\affiliation{Kyushu University, 744 Motooka, Nishi-ku, 819-0395 Fukuoka, Japan}

\author{Jeremy Atkinson\,\orcidlink{0009-0003-3287-2196}}
\affiliation{Albert Einstein Center for Fundamental Physics, Laboratory for High Energy Physics, University of Bern, Sidlerstrasse 5, CH-3012 Bern, Switzerland}

\author{Florian~U.~Bernlochner\,\orcidlink{0000-0001-8153-2719}}
\affiliation{Universit\"at Bonn, Regina-Pacis-Weg 3, D-53113 Bonn, Germany}

\author{Jianming Bian\,\orcidlink{0000-0003-3739-5424}}
\affiliation{Department of Physics and Astronomy, University of California, Irvine, CA 92697-4575, USA}

% ...

\begin{abstract}
\bigskip
The ATOMKI anomalies motivate the possibility of a new particle, the \XSeventeen, with a mass of $17~\mev$ and weak couplings to nuclei and electrons.  Such a particle may be produced at the LHC, travel 480 m in the far-forward direction, and decay to $e^+e^-$ in FASER.  In this paper, previous FASER searches for dark photons and axion-like particles (ALPs) are combined and interpreted as searches for a candidate \XSeventeen particle, the protophobic vector boson. The searches use data collected from $pp$ collisions at $\sqrt{s} = 13.6~\tev$ in LHC Run 3 from 2022 for the dark photon search and from 2022-2023 for the ALP search.  No excess events are found above the expected background, and the combined null results exclude a previously-unconstrained ``low-coupling window'' with electron coupling $6 \times 10^{-5} < \epsilon_e < 1.2 \times 10^{-4}$.  A ``high-coupling window'' remains. 
\end{abstract}
\maketitle

\blankfootnote{
\copyright~2026 CERN for the benefit of the FASER Collaboration. Reproduction of this article or parts of it is allowed as specified in the CC-BY-4.0 license.  
}

\section{Introduction \label{sec:intro}}

Light, feebly-interacting particles are a generic possibility for new physics beyond the Standard Model (BSM). In contrast to heavy particles, which can only be directly detected at the high-energy frontier, light, feebly-interacting particles may appear in a large number of experimental settings.  In particular, the decays of excited nuclei have long been considered as a way to search for new feebly-interacting particles with masses of approximately 20 MeV and below~\cite{Treiman:1978ge, Donnelly:1978ty, Savage:1986ty}.

In 2015, the ATOMKI Collaboration reported an anomaly in the decays of excited beryllium nuclei that may be explained by the existence of a new boson~\cite{Krasznahorkay:2015iga}.  An intense beam of protons was directed onto a lithium foil, creating excited beryllium nuclei ${^8}\text{Be}^*$, with a mass splitting of $18.15~\mev$ above the ${^8}\text{Be}$ ground state. Decays of ${^8}\text{Be}^* \to {^8}\text{Be}\ e^+ e^-$ were observed, and the opening angle $\theta_{e^+e^-}$ was measured. A large excess was found around $\theta_{e^+e^-} \approx 140^\circ$, providing $6.8\sigma$ evidence for a new boson with a mass of approximately 17~MeV:~the \XSeventeen or, simply, the $X$ boson.  Nuclear physics explanations have been investigated~\cite{Zhang:2017zap}, and a recent study suggests that the cosmic ray background should be studied in more detail~\cite{Benmansour:2026wxk}, but to date, no nuclear physics or experimental effect has been accepted as an explanation of the observed excess.  

The ATOMKI Collaboration subsequently studied excited helium nuclei ${^4}\text{He}^*$ at an energy 20.49 MeV above the ground state~\cite{Krasznahorkay:2021joi}.  The decays ${^4}\text{He}^* \to {^4}\text{He}\ e^+ e^-$ showed a similar large excess, but at $\theta_{e^+e^-} \approx 115^\circ$. The different locations of the $^8$Be and $^4$He excesses make them  difficult to attribute to an experimental effect, but this change is precisely what is expected for a BSM explanation in terms of a 17 MeV particle, as the larger energy gap in the $^4\text{He}$ reaction produces faster $X$ bosons and therefore more collimated decay products and smaller $\theta_{e^+e^-}$~\cite{Feng:2020mbt, Krasznahorkay:2021joi}.  

Subsequent studies have found supporting evidence for a new boson in excited $^{12}$C decays from ATOMKI~\cite{Krasznahorkay:2022pxs} and in $^8\text{Be}^*$ decays from an experiment at the VNU University of Science~\cite{Anh:2024req}. A search in the beryllium system at MEG II yielded a null result~\cite{MEGII:2024urz}, while a recent search by PADME for a resonance $e^+e^- \to X \to e^+e^-$ yielded an approximately $2\sigma$ excess for an $X$ boson mass of $17~\mev$~\cite{PADME:2025dla}. 

Shortly after the initial $^8\text{Be}$ excess was reported, the signal was explained in terms of a 17 MeV protophobic vector boson~\cite{Feng:2016jff, Feng:2016ysn}.  The required mass and couplings were later shown to also explain the location and size of the $^4\text{He}$ anomaly within theoretical and experimental uncertainties~\cite{Feng:2020mbt}.  Other possibilities have also been explored; for a review, see Ref.~\cite{Alves:2023ree}.  In brief, a new scalar particle cannot explain the $^8\text{Be}$ anomaly without violating parity~\cite{Feng:2016jff}, and a pseudoscalar~\cite{Ellwanger:2016wfe} is now disfavored~\cite{Feng:2020mbt,Barducci:2022lqd,Fieg:2026zkg}.  Axial-vector~\cite{Kozaczuk:2016nma} candidates have also been investigated, as well as the possibility of a boson with both vector and axial-vector couplings~\cite{Fieg:2026zkg}.  Their viability depends on the analysis method, including the uncertainty assigned to axial-vector nuclear matrix elements.  The multiple experimental results and proposed theoretical explanations strongly motivate further experimental studies to either confirm or exclude BSM explanations of the ATOMKI anomalies. 

One such experiment dedicated to the search for new light, feebly-interacting particles is FASER~\cite{Feng:2017uoz,FASER:2022hcn}, which operates at CERN's Large Hadron Collider (LHC). FASER is located in the far-forward direction of the LHC, approximately 480 m from the ATLAS interaction point (IP1), along the beam collision axis. From this location, the FASER main detector, consisting of veto scintillators, a tracking spectrometer, and a calorimeter, is able to detect light, long-lived particles (LLPs) that are produced at IP1, travel along the beam line, and then decay in the FASER decay volume~\cite{FASER:2018eoc}.  In this paper, two previous FASER searches, originally targeting dark photons~\cite{FASER:2023tle} and ALPs~\cite{FASER:2024bbl}, are interpreted as searches for the protophobic vector boson, and their null results are combined to yield new constraints on the $X$ boson parameter space.

\section{Protophobic Vector Boson \label{sec:protophobic}}

The $X$ boson is a massive spin-1 Abelian gauge boson that couples non-chirally to Standard Model fermions with charges $\epsilon_f$ in units of $e$.  The new Lagrangian terms are
\begin{equation}
{\cal L} = - \frac{1}{4} X_{\mu\nu} X^{\mu\nu} + \frac{1}{2} m_X^2 X_\mu X^\mu - X^{\mu} J_{\mu} \ , 
\end{equation}
where the $X$ boson has field strength $X_{\mu\nu}$ and mass $m_X$, and it couples to the current $J_{\mu} = \sum_f e \epsilon_f \bar{f} \gamma_{\mu} f$ with the fermion couplings $e \epsilon_f$. The nucleon-level current is $J^N_{\mu} = e \epsilon_p \bar{p} \gamma_{\mu} p + e \epsilon_n \bar{n} \gamma_{\mu} n$, with $\epsilon_p = 2\epsilon_u + \epsilon_d$ and $\epsilon_n = \epsilon_u + 2 \epsilon_d$.

To allow the $X$ couplings to be large enough to explain the ATOMKI anomalies without violating  stringent constraints on exotic pion decays $\pi^0 \to X \gamma$, $\epsilon_p$ must be smaller than $\epsilon_n$ by an order of magnitude~\cite{Feng:2016jff, Feng:2016ysn}.  In this work, the strict protophobic limit, where $\epsilon_p = 0$, will be assumed, and hence $\epsilon_d =  - 2 \epsilon_u$. Coherent neutrino scattering experiments further constrain the coupling of the $X$ boson to the electron neutrino to be very small, with $\epsilon_{\nu_e} < 3 \times 10^{-6}$~\cite{Denton:2023gat} with additional constraints existing for other neutrino flavors. Following Refs.~\cite{Feng:2016jff, Feng:2016ysn}, $\epsilon_{\nu} = 0$ will be assumed for all neutrinos.

Subsequent research revealed additional constraints arising from rare charged-pion decays~\cite{Hostert:2023tkg}. Avoiding them, while explaining the anomaly, requires adding both small proton and axial-vector couplings~\cite{Fieg:2026zkg}. The situation is further complicated by additional rare kaon-decay constraints~\cite{Hostert:2026xul}. In this work, a simplified vector-boson model parameterized by $\epsilon_n$ and $\epsilon_e$ will be considered. Since additional proton and axial-vector couplings have to be small, this closely approximates the phenomenology of viable full models.

To explain the $^8\text{Be}$ anomaly, $\epsilon_n = (0.8-1.4) \times 10^{-2}$~\cite{Feng:2020mbt}.  For most of this paper, a central value $\epsilon_n = 10^{-2}$ is chosen, and the electron coupling is treated as a free parameter. The conclusions are found to be insensitive to variations of $\epsilon_n$ in the favored range. The phenomenology is therefore fully specified in the $(m_X, \epsilon_e)$ plane, and the $^8\text{Be}$ anomaly is potentially resolved for $m_X=17~\mev$, subject to the constraints on $\epsilon_e$ discussed below. Results will also be presented in the $(\epsilon_n, \epsilon_e)$ plane for fixed $m_X = 17~\mev$ and for an alternative protophobic model in Appendix~\ref{sec:alternative}.

The $X$ boson decays through its electron coupling with width
\begin{equation}
\Gamma (X \to e^+ e^-) = \epsilon_e^2 \alpha \frac{m_X^2 + 2 m_e^2}{3 m_X} 
\sqrt{1 - \frac{4 m_e^2}{m_X^2}} \ ,
\label{eq:width}
\end{equation}
where $\alpha$ is the fine structure constant, and $m_e$ is the mass of the electron. The loop-mediated decay $X \to \gamma \gamma \gamma$ is negligible for couplings of interest. 

To explain the ATOMKI results, the $X$ boson must decay promptly in the ATOMKI experiments so that the $e^+ e^-$ decay products are detected and the $\theta_{e^+ e^-}$ measurements are not distorted, leading to the bound $\epsilon_e \agt 1.3 \times 10^{-5}$~\cite{Feng:2016jff, Feng:2016ysn}. Constraints from many other experiments, including E141~\cite{Riordan:1987aw}, E137~\cite{Bjorken:1988as}, Orsay~\cite{Davier:1989wz}, KLOE~\cite{Anastasi:2015qla}, NA64~\cite{NA64:2019auh}, BaBar~\cite{BaBar:2014zli}, and PADME~\cite{PADME:2025dla} 
(see \cref{fig:faser_lmits_x17}) now limit the acceptable range to two windows: 
\begin{align}
\text{High-coupling: } & 6 \times 10^{-4} < \epsilon_e < 1.5 \times 10^{-3} \\
\text{Low-coupling: } & 6 \times 10^{-5} < \epsilon_e < 1.2 \times 10^{-4} \ .
\end{align}

Note that the low-coupling window was previously often considered excluded by E141~\cite{Riordan:1987aw}, based on a 2012 overview of past electron beam dump experiments~\cite{Andreas:2012mt}.  This overview included a reinterpretation of the axion search results of E141. Recently, a dedicated re-analysis of the E141 results~\cite{Celentano:2026gkd} concluded that more detailed considerations weaken the E141 bound, opening up the low-coupling window.  It is the low-coupling window that is probed by the FASER searches described here.

\section{Forward \XSeventeen Production at the LHC \label{sec:signal}}

At first sight, since the $X$ boson is both protophobic and pion-phobic, it would appear to be a difficult target for FASER, as past searches have relied on the targeted particles being produced in pion decays, which is not possible for the $X$ boson, or being produced through dark bremsstrahlung from protons, which would also seem to be impossible. In fact, however, the $X$ boson can be produced through dark bremsstrahlung, as well as in $\eta$ decays. 

The forward production of $\eta$ mesons is modeled using \texttt{EPOS-LHC}~\cite{Pierog:2013ria} as implemented in \text{CRMC}~\cite{CRMC}. The associated flux uncertainty is estimated by considering the envelope of predictions provided by \texttt{EPOS-LHC}, \texttt{SIBYLL~2.3d}~\cite{Riehn:2019jet}, and  \texttt{QGSJET~2.04}~\cite{Ostapchenko:2010vb}. The branching fraction for $\eta \to X \gamma$ decays is~\cite{Ilten:2018crw} 
\begin{equation}
B(\eta \to X \gamma) \simeq 0.5 \, \epsilon_n^2 \, B(\eta \to \gamma\gamma) \, \left(1-m_X^2/m_\eta^2\right)^3 \, .
\label{eq:br_eta}
\end{equation}
The spin correlations between production and decay have been shown to have a negligible effect for FASER searches~\cite{Feng:2025gji, AristizabalSierra:2026jsm} and are therefore neglected. $X$ boson production via heavier meson decays is found to be sub-leading and is therefore also not considered.

The differential cross section for $X$ production in proton bremsstrahlung can be written as~\cite{Feng:2017uoz, Foroughi-Abari:2021zbm}
\begin{equation}
\frac{d\sigma}{dz dp_T^2} =  \epsilon_n^2 \,   \sigma^{nsd}_{pp}(s') \, w(z,p_T^2) \, \left| F(m_{X}^2) \right|^2 \, f(p') \ ,
\label{eq:xs_brem}
\end{equation}
where $z =  p_{X}/p_{\rm beam}$ is the momentum fraction transferred to the $X$ boson, and $p_T$ is its momentum transverse to the beam. The non-single diffractive cross section $\sigma^{nsd}_{pp}(s')$ at $s' \approx (1-z)s$ with $s = (13.6~\text{TeV})^2$ is taken from Ref.~\cite{Likhoded:2010pc}, as recommended in Ref.~\cite{Foroughi-Abari:2021zbm}. The splitting functions $w$ are those derived via the quasi-real approximation including the Dawson correction, as presented in Ref.~\cite{Foroughi-Abari:2024xlj}, and the form factors $F(m_X^2)$ are those derived for arbitrary charge assignments, as presented in Ref.~\cite{Kling:2025udr}.

\begin{figure*}[t]
\includegraphics[width=0.9\textwidth]{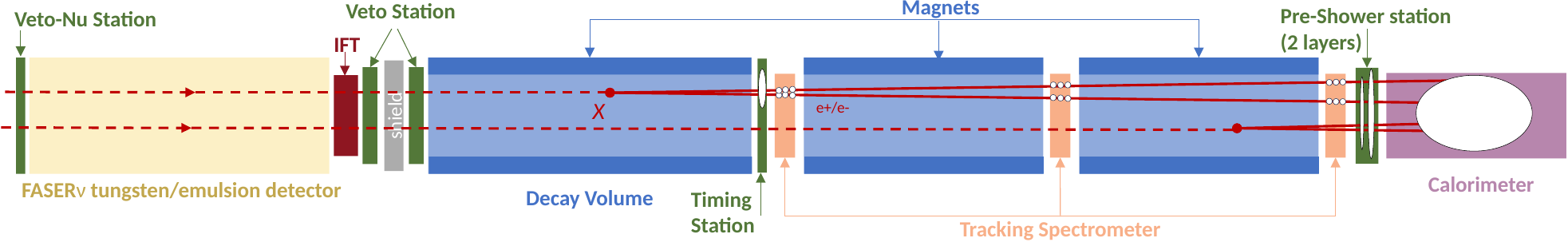}
\caption{A lateral schematic view of the FASER detector, showing the different detector systems, as well as the signature of an $X$ boson decaying to an $e^+ e^-$ pair. If the decay occurs within the decay volume, the resulting $e^+e^-$ produces a pair of charged tracks in the spectrometer, making the event a candidate for the dark photon search. Alternatively, if the decay occurs downstream of the decay volume, the event becomes a candidate for the axion-like particle (ALP) search. The utilization of the timing station as an additional veto scintillator in the ALP search ensures orthogonality between the two searches.}
    \label{fig:x17_faser_schema}
\end{figure*}

Both the Dirac form factor $F_1(t)$ and the Pauli form factor $F_2(t)$ are accounted for and evaluated at $t=m_{X}^2$. The interference term is included, so that $w |F|^2 = w_1 |F_1|^2 + w_2 |F_2|^2 + w_{1,2} \text{Re}(F_1 F_2^*)$. This also explains why $X$ can be produced in proton bremsstrahlung:~although $F_1(0)=0$ by construction, both form factors are non-zero for $m_X>0$ and, therefore, allow efficient production through proton bremsstrahlung. The form factor \texttt{Model~1} is used, and the associated uncertainty is estimated from the envelope of predictions provided by \texttt{Models~1}, \texttt{2}, and \texttt{3} of Ref.~\cite{Kling:2025udr}. Although this result assumes $\epsilon_s = \epsilon_d$, the result is independent of $\epsilon_s$ for $m_X \ll m_\phi$, where $\epsilon_s$ is the coupling to the strange quark, and $m_\phi$ is the mass of the $\phi$ meson.

The validity of coherent bremsstrahlung is ensured by requiring the intermediate proton to stay intact, and its momentum $p' = p_{\rm beam} \!-\! p_{X}$ is prevented from becoming  highly off-shell by applying a damping factor~\cite{Foroughi-Abari:2021zbm}
\be
f(p')=\frac{\Lambda_p^4}{\Lambda_p^4 + (p'^2-m_p^2)^2} \ .
\ee
Here $\Lambda_p = 2 \pi \Lambda_\text{QCD}\sim 1.5~\gev$, and the uncertainty is estimated by varying $\Lambda_p$ between $1.0~\gev $ and $2.0~\gev$~\cite{Foroughi-Abari:2024xlj}.

The production of $X$ bosons via bremsstrahlung off neutrons colliding with the TAN, the absorber for neutral particles located about 150~m downstream from IP1, is included. This is simulated, following the description of Ref.~\cite{Kling:2025udr}, with neutron fluxes from \texttt{EPOS-LHC} and the modeling of bremsstrahlung discussed above with the appropriate form factors for neutrons. The uncertainty in the neutron flux is taken to be correlated with the uncertainty in the $\eta$-meson flux.

\section{FASER Detector \label{sec:faser}}

The FASER detector is located approximately $480~\m$ downstream of IP1 in the TI12 tunnel that used to connect the LHC with the Super Proton Synchrotron (SPS). Although the detector is roughly aligned with the beam collision axis line of sight (LOS) at IP1, there is a small offset induced by the crossing angle at IP1. In 2022-2023 the crossing angle at IP1 was $-160~\micrad$, resulting in a vertical downward shift of the LOS of $6.5~\cm$ with respect to the center of the detector. With an active transverse radius of $10~\cm$, FASER covers an angular acceptance corresponding to a pseudorapidity of $\eta>9.2$.

The detector is described in detail in Ref.~\cite{FASER:2022hcn}; a simplified schematic is shown in \cref{fig:x17_faser_schema}.  The detector includes the Veto-Nu station, the FASER$\nu$ emulsion detector, the interface tracker (IFT), the veto station, the decay volume, the timing station, the tracking spectrometer, the preshower scintillators, and the electromagnetic (EM) calorimeter.

For this paper, the FASER$\nu$ emulsion detector is not used, but the eight interaction lengths of tungsten do provide some passive shielding against potential backgrounds. The two veto stations provide highly-efficient rejection of incoming muons, since their minimum transverse size of $30\times30~\cm^2$ is significantly larger than the active region of the detector. The decay volume and tracking spectrometer are immersed in a $0.57~\text{T}$ magnetic field, bending charged-particle trajectories in the vertical direction. The timing scintillator layer aids in triggering and timing measurements, and it can also function as an additional veto layer for decays after the decay volume.

The tracking spectrometer consists of three tracking stations, each with three layers of double-sided silicon microstrip detectors, using spare Semi-Conductor Tracker (SCT) barrel modules from ATLAS~\cite{Abdesselam:2006wt}. The $40~\mathrm{mrad}$ stereo angle between the two sides allows the layer to have a hit resolution of $20~\micm$ ($0.6~\mm)$ in the bending (non-bending) direction of the magnetic field. Downstream of the tracker is the preshower detector, consisting of two scintillator layers each, preceded by a $3~\mm$-thick layer of tungsten. To prevent backsplash into or from the preshower, two $5~\cm$-thick graphite blocks are placed at the front and back ends of the preshower detector. 

Finally, the most downstream component of the detector is the EM calorimeter. The calorimeter is composed of four modules in a $2 \times 2$ array in the transverse plane. These are from the outer EM calorimeter of the LHCb detector~\cite{LHCb:2000vji}, and each module has a total depth of 25 radiation lengths and is $12 \times 12~\mathrm{cm}^2$ in the transverse plane. The energy resolution has been measured to be ${\cal O}(1\%)$ in the high-energy range most relevant for this analysis, using data collected in a test beam study at the CERN SPS carried out in July 2021~\cite{FASER:2022hcn}.

\begin{table*}
    \centering
        \begin{tabular}{cccc}
            \hline\hline
            \textbf{Source}                 &
            \textbf{Dark Photon}            &
            \ \ \ \textbf{ALP}\ \ \        &
            \textbf{Comment}                                                  \\
            \hline
            Luminosity
                                            & 2.2\%
                                            & 2.0\%
                                            & 2.2\% (conservative)         \\
            Calo-Energy Scale
                                            & 6.0\%
                                            & 6.0\%
                                            & Correlated between searches                  \\
            Single-Track Efficiency
                                            & 3.0\%
                                            & -
                                            & Only for dark photon search             \\
            Two-Track Efficiency
                                            & 7.0\%
                                            & -
                                            & Only for dark photon search             \\
            Track Momentum Scale
                                            & 5.0\%
                                            & -
                                            & Negligible                \\
            Track Momentum Resolution
                                            & 5.0\%
                                            & -
                                            & Negligible                       \\
            PS Ratio
                                            & -
                                            & 13\%
                                            & Only for ALP search        \\
            PS-1 nMIP
                                            & -
                                            & 20\%
                                            & Negligible on $e^+e^-$ signals \\
            \hline \hline
        \end{tabular}
    \caption{Sources of experimental systematic uncertainties in the dark photon and ALP searches. The luminosity and calorimeter energy scale uncertainties are common to both searches and are treated as fully correlated.}
    \label{tab:systematics}
\end{table*}

\section{FASER Searches \label{sec:searches}}

FASER is designed, in part, to search for light, feebly-interacting particles produced at the LHC.  FASER has previously reported world-leading searches for dark photons~\cite{FASER:2023tle} and for ALPs and other LLPs with photonic signatures~\cite{FASER:2024bbl}. 

This paper combines the dark photon search from Ref.~\cite{FASER:2023tle} and the ALP search from Ref.~\cite{FASER:2024bbl}, and interprets this combination for the $X$ boson. The combination of these searches enables the entire magnetized region of the detector to be utilized as a viable decay volume for the $X$ boson, as illustrated in \cref{fig:x17_faser_schema}.

\subsubsection*{FASER's Dark Photon Search}

FASER searched for dark photons in $27~\ifb$ of data collected in 2022 during Run 3 of $pp$ collisions at $\sqrt{s}=13.6~\mathrm{TeV}$~\cite{FASER:2023tle}. The search targeted dark photons decaying into $e^+ e^-$ pairs within the decay volume. The incoming dark photon does not leave a signal in any of the upstream veto scintillators, while the decay products produce a signal consistent with two minimum-ionizing particles (MIPs) in the timing scintillator. The electrons are subsequently reconstructed by the tracking spectrometer as a pair of tracks with a large energy deposit of at least $500~\gev$ in the calorimeter.

Various sources of backgrounds were considered for this search.  Most backgrounds were found to be negligible with tiny contributions from neutral hadrons and neutrinos. In particular, the neutrino background was strongly suppressed by the stringent tracking requirements, resulting in an expected contribution of \mbox{$(1.5 \pm 1.95~(\mathrm{tot.})) \times 10^{-3}$} events. 

FASER observed no events in this signal region, consistent with the total expected background of \mbox{$(2.3 \pm 2.3~(\mathrm{tot.})) \times10^{-3}$} events.

\subsubsection*{FASER's ALP Search}

The ALP search utilized $57.7~\ifb$ of data, comprising the dataset used in the dark photon search, as well as additional data collected in 2023~\cite{FASER:2024bbl}.

This search targeted LLPs decaying into photonic or EM final states. It utilized the timing scintillator as an additional veto layer, since the electrically-neutral ALPs or photons do not deposit any signal in the scintillators. This allowed the ALP search to be orthogonal to the dark photon search, while still being sensitive to LLPs decaying into electrons after the timing scintillator. The photons or electrons are detected through the evidence of a shower in the preshower detector accompanied by a large energy deposit of at least $1.5~\tev$ in the calorimeter.  The evidence of a shower is quantified using the ratio of charge deposited in two scintillator planes of the preshower, where a developing shower would have a higher charge deposit in the second layer leading to a higher preshower ratio.

The backgrounds considered are similar to those in the dark photon search. Although most remain negligible, the neutrino background is larger than in the dark photon search, primarily due to $\nu_e$ charged-current interactions in the downstream detector components, particularly the preshower and calorimeter, which can mimic the signal signature. 

FASER observed one event in the signal region, consistent with the expected neutrino background of \mbox{$0.44\pm0.39~(\mathrm{tot.})$} events.

\begin{figure*}[th]
\includegraphics[width=0.49\linewidth]{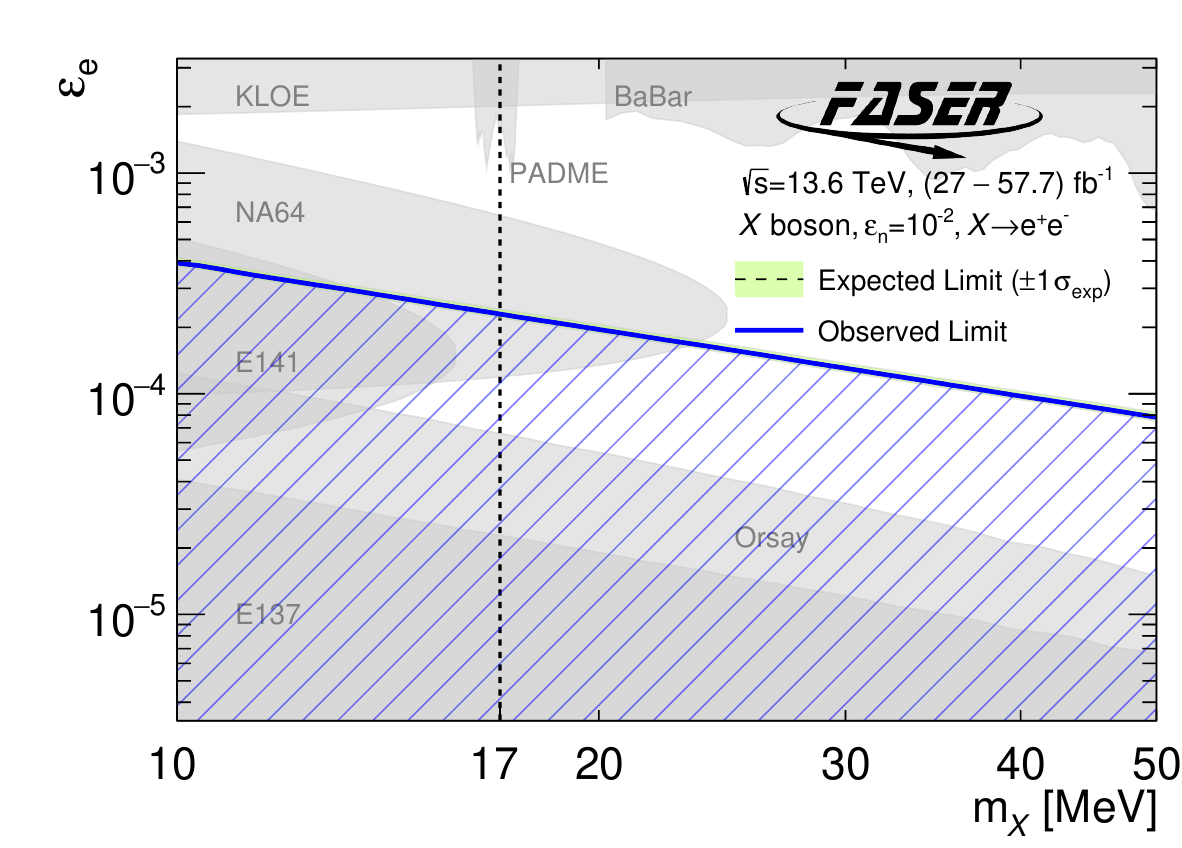}
\hfill
\includegraphics[width=0.49\linewidth]{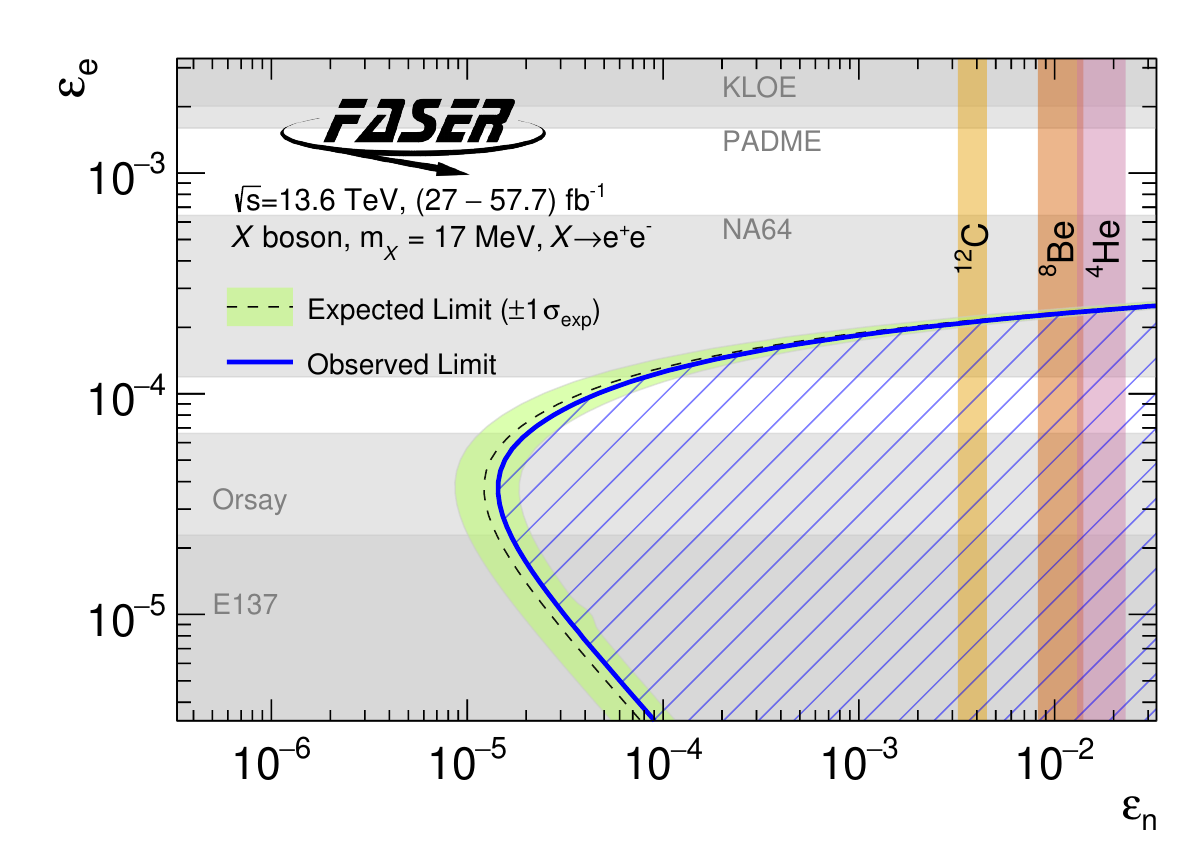}
\caption{90\% confidence level exclusions of the $X$ boson, using a statistical combination of the FASER dark photon (27 $\ifb$)~\cite{FASER:2023tle} and ALP (57.7 $\ifb$)~\cite{FASER:2024bbl} searches. {\bf Left:} Excluded region in the $(m_X, \epsilon_e)$ plane where $\epsilon_n = 10^{-2}$ is fixed to explain the $^8\text{Be}$ ATOMKI anomaly. {\bf Right:} Excluded region in the $(\epsilon_e, \epsilon_n)$ plane for the fixed mass $m_X = 17~\mev$. The ranges of $\epsilon_n$ that explain the $^{12}\text{C}$, $^8\text{Be}$, and $^4\text{He}$ ATOMKI anomalies are shown as vertical colored bands.  In both panels, the region excluded by FASER is indicted by the hatches and regions excluded by other experiments are shown in gray~\cite{Anastasi:2015qla, BaBar:2014zli,Bjorken:1988as, PADME:2025dla, NA64:2019auh, Celentano:2026gkd, Davier:1989wz}.}
\label{fig:faser_lmits_x17}
\end{figure*}

\subsubsection*{Systematic Uncertainties}

The two searches considered several sources of systematic uncertainty. The sources of systematic uncertainties in the signal production are described in \cref{sec:signal}.  The uncertainties associated with the parent-hadron flux are estimated by varying the underlying flux model. These uncertainties are correlated across the production channels. The uncertainty associated with the damping factor was found to be dominant in the region of interest, accounting for approximately  a 40\%  uncertainty on the total signal production.

The dominant experimental uncertainties arise from the modeling of the detector response in the Monte-Carlo simulation. 

For the dark photon search, the main systematic uncertainties are associated with track reconstruction, with uncertainties of $3\%$ for single-track finding and $7\%$ for two-track finding. There are also a 5\% uncertainties associated with both the track momentum scale and resolution; however, these uncertainties were found to have a negligible impact on the signal yield.
In contrast, the ALP search was primarily affected by uncertainties in the preshower response, with a $13\%$ uncertainty in the preshower ratio and a $20\%$ uncertainty in the MIP response of the second preshower layer. The latter uncertainty was found to be negligible for $e^+e^-$ signals. 

Since the underlying searches have different selection requirements, the systematic uncertainties of one may not be applicable to the other, and in these cases, the uncertainties are treated as uncorrelated. The sources of uncertainty that are common to both searches are from the calorimeter and the luminosity, which are taken to be fully correlated between the searches. A $6\%$ uncertainty in the calorimeter energy scale is shared between the two searches. The resulting effect on the signal yield is evaluated by shifting the corresponding parameter values used to define the signal regions.

The integrated luminosity uncertainty for the dark photon search was $2.2\%$, as provided by ATLAS~\cite{ATLAS:2023hgc}. With the increased dataset used in the ALP search, this uncertainty was reduced to $2.0\%$~\cite{ATLAS:2024tzv}. For this analysis, a conservative uncertainty of $2.2\%$ is adopted.

A summary of the experimental systematic uncertainties is given in \cref{tab:systematics}.

\section{Results \label{sec:result}}

Signal events are generated using \texttt{FORESEE}~\cite{Kling:2021fwx}, considering $X$-production via $\eta$ decays, proton bremsstrahlung in collision and neutron bremsstrahlung at the TAN. Proton bremsstrahlung was found to be the main production mode. The signal yields are evaluated using a parameterization of the signal efficiency determined from Monte Carlo simulations. For the ALP selection, the parameterization given in Appendix C of Ref.~\cite{FASER:2024bbl} is used. The dark photon search does not provide an explicit parameterization; however, following a similar strategy to that used for the ALP selection, a two-dimensional parameterization in energy and radius has been developed. Details of this parameterization can be found in Appendix~\ref{sec:param_signal_eff}.

The resulting parameterizations are applied to the generated signal events to evaluate the corresponding yields in each signal region. The systematic uncertainties are also evaluated by the same procedure, wherein the signal region definition is varied by the corresponding uncertainty as defined in the previous section. This procedure allows both the signal yields and their associated systematic uncertainties to be determined consistently.

A statistical interpretation of the result is performed using a profile likelihood approach using the \texttt{HistFitter} statistical analysis framework~\cite{Baak:2014wma}. The $\mathrm{CL}_s$ method~\cite{Read:2002hq} is used to evaluate the exclusions at the 90\% confidence level. 

The resulting exclusion in the $(m_X, \epsilon_e)$ plane is shown in the left panel of \cref{fig:faser_lmits_x17}. Since $\epsilon_n$ is taken to be $10^{-2}$, the production rate is fixed. The reach is limited at high couplings $\epsilon_e$ by the decreasing $X$ boson lifetime; for $\epsilon_e > 3\times 10^{-4}$, a 17 MeV $X$ boson is not sufficiently long-lived to reach FASER. At low couplings $\epsilon_e$, the lifetime becomes very long. The reach is then limited by the probability that it decays inside FASER and extends down to $\epsilon_e \approx 2\times 10^{-8}$ at $m_X = 17~\mev$. 

The right panel of \cref{fig:faser_lmits_x17} shows the reach of FASER in the $(\epsilon_n, \epsilon_e)$ plane for fixed $m_X = 17~\mev$. The ranges of $\epsilon_n$ that explain the ATOMKI anomalies are shown as vertical colored bands.  The width of these bands is determined by varying the $X$ boson decay widths within the $1\sigma$ best-fit ranges determined by experiment and by considering theoretical uncertainties from isospin mixing of the excited nuclei with nearby states, isospin breaking in the EM transition operators, and varying $\epsilon_p$ within the experimentally-allowed range~\cite{Feng:2020mbt}.  These bands could also be broadened by uncertainties in the size of backgrounds from Standard Model internal pair conversion processes at ATOMKI. Nevertheless, given the sensitivity of FASER, for a potential $17~\mathrm{MeV}$ vector boson, this analysis completely excludes the low-coupling window, leaving only the high-coupling window as a viable region of parameter space.

\section{Summary \label{sec:summary}}

The ATOMKI anomalies are some of the most tantalizing anomalies in particle physics. A protophobic vector boson has been proposed as a possible explanation of the observed excess, with the current experimental constraints leaving two regions of parameter space with potentially viable couplings.  This work shows that the combination of FASER dark photon and ALP searches excludes electron couplings in the range $2\times 10^{-8} < \epsilon_e < 3\times 10^{-4}$ for a protophobic vector boson of mass $m_X = 17~\mev$. Consequently, the low-coupling window for a protophobic explanation of the \XSeventeen is excluded by existing FASER data, leaving only the high-coupling window as a viable possibility. This result highlights the capability of FASER and other far-forward LHC experiments to search for LLPs and provide complementary and sensitive tests of BSM explanations of anomalies in low-energy experiments.

\section*{Acknowledgments}
\label{sec:Acknowledgments}

We thank CERN for the excellent performance of the LHC and the technical and administrative staff members at all FASER institutions for their contributions to the success of the FASER experiment. We thank the ATLAS Collaboration for providing us with accurate luminosity estimates for the Run 3 LHC $pp$ collision data. We also thank the CERN STI group for providing detailed \texttt{FLUKA} simulations of the muon fluence along the LOS.

This work was supported in part by Heising-Simons Foundation Grant Nos.~2019-1179 and 2020-1840, Simons Foundation Grant No.~623683, JSPS KAKENHI Grant Nos.~19H01909, 22H01233, 20K23373, 23H00103, 20H01919, and 21H00082, the joint research program of the Institute of Materials and Systems for Sustainability, ERC Consolidator Grant No.~101002690, DFG grant~SCHO~1527/13-1 and KL~3880/1-1, Royal Society Grant No.~URF$\backslash$R1$\backslash$201519, UK Science and Technology Funding Councils Grant No.~ST/ T505870/1, Leverhulme Trust Grant No.~RPG-2026-156, the National Natural Science Foundation of China, Tsinghua University Initiative Scientific Research Program, and the Swiss National Science Foundation.

\appendix


\section{Parameterized Signal Efficiencies}
\label{sec:param_signal_eff}

The parameterized signal efficiencies are evaluated for the dark photon selection using a dedicated high-statistics sample of LLPs decaying into the $e^+e^-$ final state. The samples are generated using \texttt{FORESEE}~\cite{Kling:2021fwx} and are simulated using FASER's dedicated \texttt{GEANT4}~\cite{GEANT4:2002zbu} simulation, which incorporates a detailed description of the detector geometry. The reconstruction uses the same pipeline as the data. The efficiency of the LLP passing the dark photon search's selections is evaluated as a function of the LLP energy ($E$) and the radial position ($R$) of the decay vertex. The fit function used is 
\begin{equation}
    \varepsilon\left(E, R \right) = 
    \varepsilon_{\mathrm{max}} \
    \frac
    {\left[ 1 + \tanh(-\left( \frac{R - R_c}{\Delta R}\right)\right]/2}
    {1 + \exp\left(-\frac{E -E_c}{\Delta E} \right)} \ ,
    \label{eq:fit}
\end{equation}
where $\varepsilon_{\mathrm{max}}$, $E_c$, $R_c$, $\Delta E$, and $\Delta R$ are parameters determined from the fit and are listed in \cref{tab:fit_params}. The projections of the two-dimensional fits onto the energy of the LLP is shown in \cref{fig:fits}. The dark photon search requires a minimum energy of $500~\gev$ in the calorimeter, which is reflected in the ``turn-on'' behavior of the efficiency curve at the same energy. The efficiencies for calorimeter energy cuts of 470~GeV and 530~GeV are used to estimate the systematic uncertainty corresponding to the 6\% calorimeter energy resolution.

\begin{table*}
    \centering
    \setlength{\tabcolsep}{7pt} 
    \begin{tabular}{cccccc}
    \hline \hline
         Calorimeter EM Energy Cut & $\varepsilon_{\mathrm{max}}$ & $E_c$~[GeV] & $\Delta E$~[GeV] & $R_c~[\mm]$ & $\Delta R~[\mm]$  \\
    \hline
         $500$~GeV & 0.242 & 503 & 6.57 & 91.7 & 2.6 \\
         $470$~GeV & 0.241 & 465 & 2.04 & 91.7 & 2.6 \\
         $530$~GeV & 0.241 & 532 & 6.87 & 91.8 & 2.5 \\
     \hline \hline
    \end{tabular}
    \caption{Fit parameters of the simultaneous fit using \cref{eq:fit} to the LLP energy and radial position of the decay vertex for the dark photon selection.}
    \label{tab:fit_params}
\end{table*}

\begin{figure*}[th]
\centering
\includegraphics[width=0.85\linewidth]{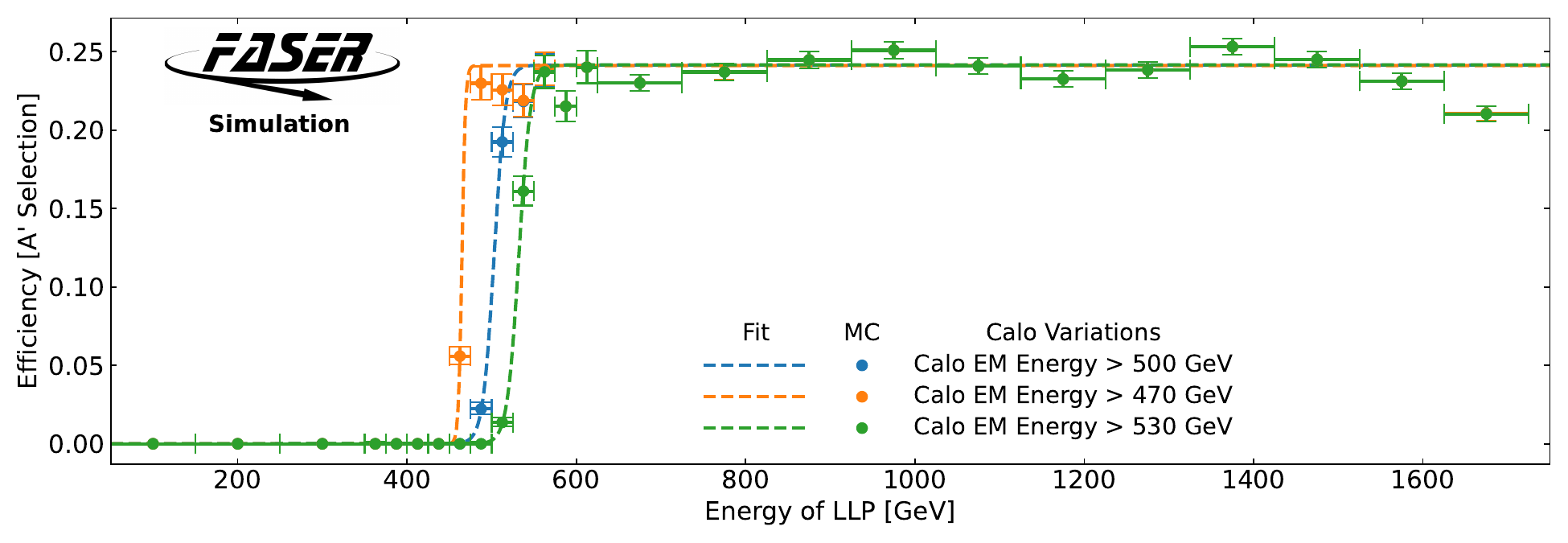}
\caption{Parameterized signal efficiencies for the dark photon selection for LLPs decaying inside FASER as a function of the underlying energy of the LLP. For the one-dimensional energy fit, only events with \mbox{$R<90~\mm$} are shown. This restriction ensure that the plots are not affected by sampling artifacts in the other variable. The fit parameters listed in \cref{tab:fit_params} are obtained from the simultaneous fit to both variables. The fits corresponding to the varied signal regions used to account for the systematic uncertainty in the calorimeter energy are shown in orange and green.}
\label{fig:fits}
\end{figure*}

\section{Alternative Protophobic Model}
\label{sec:alternative}

The main part of the paper considers the protophobic gauge boson with a fixed value of the neutron coupling $\epsilon_n = 10^{-2}$ chosen to explain the ATOMKI $^8$Be anomaly. An alternative realization of the model, in which  non-vanishing couplings are assumed to be proportional to one gauge coupling $\epsilon \equiv \epsilon_n = \epsilon_e$, was considered in Refs.~\cite{Ilten:2018crw, Kling:2025udr}. This model, by construction, respects the SINDRUM~\cite{SINDRUM:1989qan} constraint on $\pi^+ \to e^+ \nu X$, which requires $\epsilon_e - \epsilon_n < 8.5\times 10^{-5}$~\cite{Hostert:2023tkg}. Its phenomenology is completely specified in the $(m_X, \epsilon)$ plane.

The region of parameter space constrained by FASER is shown in \cref{fig:x17_b}. As for the protophobic gauge boson with fixed neutron couplings, the reach at high $\epsilon$ is limited by the decreasing $X$ boson lifetime. However, the signal production rate also scales proportional to $\epsilon^2$, leading to a lower bound of the FASER contour around $\epsilon = 2 \times 10^{-5}$. FASER excludes the protophobic model for $m=17$~MeV and couplings in the range $2\times 10^{-5} < \epsilon < 1.5\times 10^{-4}$.

\begin{figure}
\includegraphics[width=\linewidth]{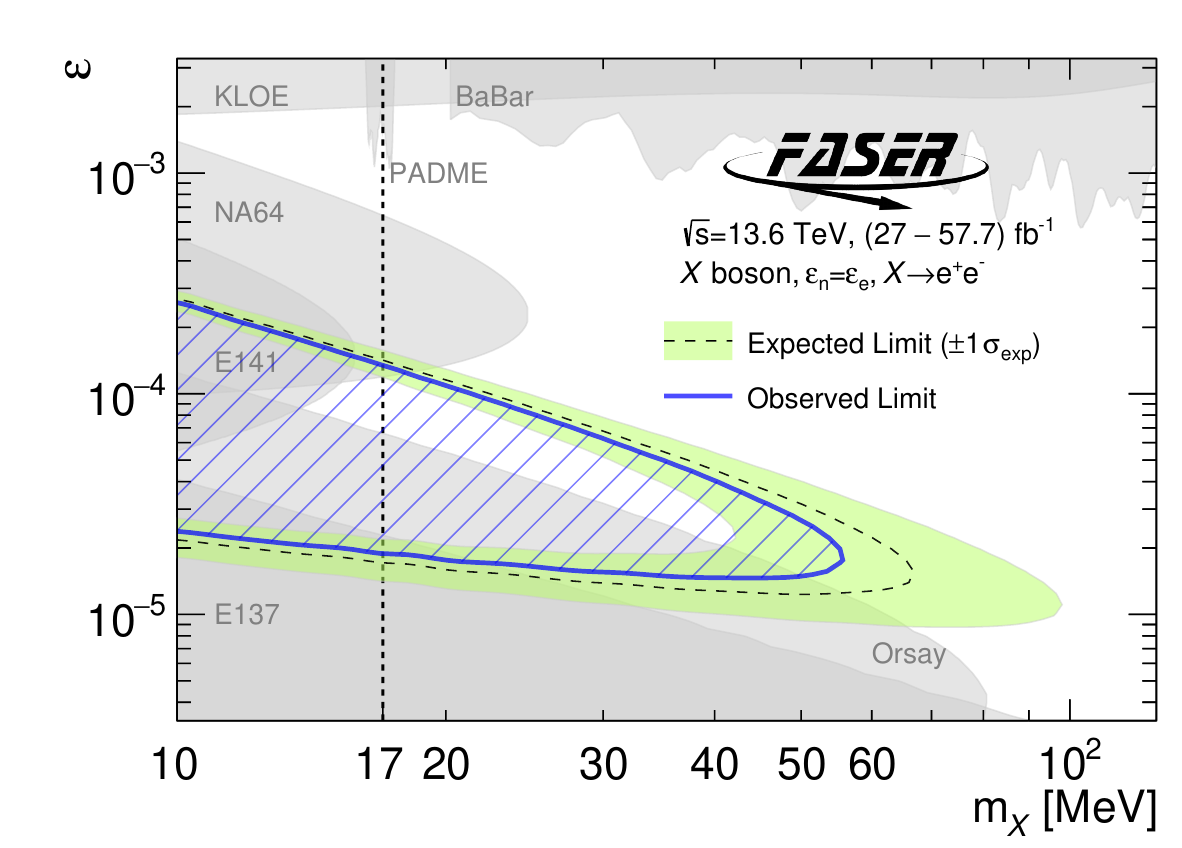}
\caption{90\% confidence level exclusion of $X$ boson, using a statistical combination of FASER dark photon ($27~\ifb$)~\cite{FASER:2023tle} and ALP ($57.7~\ifb$)~\cite{FASER:2024bbl} searches.  This exclusion region is defined in the $(m_X, \epsilon)$ plane where the couplings are taken to be proportional $\epsilon = \epsilon_n = \epsilon_e$.  Regions previously excluded by other experiments are shown in gray~\cite{Anastasi:2015qla, BaBar:2014zli,Bjorken:1988as, PADME:2025dla, NA64:2019auh, Celentano:2026gkd, Davier:1989wz}.
}
\label{fig:x17_b}
\end{figure}

\bibliographystyle{utphys}
\bibliography{referencesX17}

\end{document}